\documentclass[twocolumn,superscriptaddress,showpacs,amsmath,amssymb,prl]{revtex4}
\pdfoutput=1
\usepackage{booktabs,graphicx,pstricks,verbatim,url}
\usepackage{graphicx}
\usepackage{dcolumn}
\usepackage{bm}

\def\eeq{\end{equation}}
\def\beq{\begin{equation}}
\def\eeq{\end{equation}}
\def\ba{\begin{eqnarray}}
\def\aa{\end{eqnarray}}
\newcommand{\sun}{\odot}
\newcommand{\be}{\begin{eqnarray}}
\newcommand{\ee}{\end{eqnarray}}
\newcommand{\lsim}{\stackrel{<}{\sim}}
\newcommand{\gsim}{\stackrel{>}{\sim}}

\newcommand{\gev}{{\rm GeV}}

\newcommand{\mev}{{\rm MeV}}

\newcommand{\vev}[1]{{\langle #1 \rangle}} 
\newcommand{\disc}[1]{} 

\begin{document}
\input{epsf}

\title{Cores in Dwarf Galaxies from Dark Matter with a Yukawa Potential}

\author{Abraham Loeb} 
\affiliation{Institute for Theory \&
Computation, Harvard University, 60 Garden St., Cambridge, MA 02138}
\author{Neal Weiner} 
\affiliation{Center for Cosmology and Particle
Physics, Department of Physics, New York University, New York, NY
10003} 
\affiliation{School of Natural Sciences, Institute for
Advanced Study, Princeton, NJ 08540}

\begin{abstract} 

We show that cold dark matter particles interacting through a Yukawa
potential could naturally explain the recently observed cores in dwarf
galaxies without affecting the dynamics of objects with a much larger
velocity dispersion, such as clusters of galaxies. The velocity
dependence of the associated cross-section as well as the possible
exothermic nature of the interaction alleviates earlier concerns about
strongly interacting dark matter. Dark matter evaporation in low-mass
objects might explain the observed deficit of satellite galaxies in
the Milky Way halo and have important implications for the first
galaxies and reionization.

\end{abstract}

\pacs{95.35+d,98.80-k,98.62-g}
\date{\today}
\maketitle

\paragraph*{Introduction.}
The collisionless cold dark matter (CDM) model has been highly
successful in accounting for the gravitational growth of density
perturbations from their small observed amplitude at early cosmic
times (as imprinted on the cosmic microwave background anisotropies
\cite{WMAP}) to the present-day structure of the Universe on large
scales. However, it is far from clear that the predictions of this
model are valid on small scales.

New data on low mass galaxies indicate that their dark matter
distribution has a core \cite{Blok}, in contrast to the cusped profile
expected from collisionless CDM simulations \cite{NFW}.  The mean
value of the inner logarithmic slope of the mass density profile in
seven dwarf galaxies within the THINGS survey is observed to be
$-0.29\pm 0.07$ \cite{Oh}, much shallower than the expected slope of
$\sim -1$ from pure CDM simulations.  Moreover, the dynamics of dwarf
spheroidal galaxies, such as Fornax \cite{Goerdt}, Ursa-Minor
\cite{Kleyna}, and Sculptor \cite{Walker}, whose luminosities and
dynamical masses are smaller by 2-3 orders of magnitude than the
THINGS galaxies, indicates a characteristic core density of $\sim
0.1\pm 0.05 M_\odot~{\rm pc^{-3}}= (7\pm 4) \times 10^{-24}~{\rm
g~cm^{-3}}$.  Since these dwarf spheroidals are dominated by dark
matter throughout, it is challenging to explain their inferred cores
by the gravitational interaction of the dark matter with the baryons
\cite{Gnedin}.  Although it is conceivable that powerful gas outflows
from an early baryon-dominated nucleus would reduce the central dark
matter density in luminous galaxies \cite{Gov,Oh:2010mc}, the
formation of a massive baryonic nucleus would initially compress the
CDM \cite{Blum} and exacerbate the discrepancy that needs to be
resolved \cite{Gnedin}, and also potentially violate the observed low
luminosities from dwarf galaxies at higher redshifts
\cite{White,Guo}. High-redshift observations of dwarf galaxies must
find evidence for the required strong feedback phase, or else an
alternative process is at play.  Some recent simulations that include
feedback do not observe the appearance of cores within the lowest
luminosity galaxies \cite{2010MNRAS.402.1599S}.

To alleviate early signs of the above discrepancy, Spergel \&
Steinhardt \cite{SIDM} adopted the Strongly-Interacting Dark Matter
(SIDM) model \cite{Carlson,machecek,laix} in which the dark matter has
a large cross-section for self interaction. It was expected that if
dark matter scatters in the cores of galaxies, then it might resemble
a fluid with a flatter central density profile. The SIDM proposal fell
out of favor because: {\it (i)} gravitational lensing and X-ray data
indicate that the cores of clusters of galaxies are dense and
ellipsoidal, whereas SIDM would predict them to be shallow and
spherical \cite{Miralda,Yoshida}; {\it (ii)} relaxation of the core
ultimately generates an even denser nucleus after many collision times
due to the gravothermal catastrophe, familiar from core collapse of
globular clusters \cite{Balberg}, although this evolution might take
more than the Hubble time; {\it (iii)} dwarf galaxies would be
expected to evaporate when interacting with the higher velocity
particles of their host halo; and {\it (iv)} theoretical biases
suggested that the required cross section was incompatible with
popular models of Weakly-Interacting Massive Particles (WIMPs).

Recently, there has been growing interest in the possibility that
WIMPs exhibit ``dark forces'' as a means to address a wide range of
anomalies \cite{ArkaniHamed:2008qn}.  In particular, it was realized
that a new force carrier $\phi$ (scalar or vector) might
naturally mediate a long-range interaction on the scale of the
de-Broglie wavelength of the WIMPs, leading to a self-interaction
cross-section for scattering that is much greater than for WIMP
annihilation. The studied forces have a variety of scales in them,
from the screening scale set by the mass of the carrier particle
$m_\phi$ to the non-perturbative scale set by its coupling, $\alpha_d
m_\phi$.  Moreover, these forces are naturally accompanied by new
energy states.  Up- and down-scattering processes (which are endo- and
exo-thermic, respectively) naturally introduce yet another scale into
the problem, set by the energy splitting $\delta$ between the
states. While similar models have been previously considered
\cite{Hogan:2000bv}, our regime of interest was assumed to be
unsuitable due to WIMP capture and WIMP annihilation. As it turns out,
these processes are generally weaker than scattering and so this range
of parameters remains open. Nonetheless, the absence of dramatic
departures from CDM predictions has allowed important constraints to
be placed \cite{Feng:2009hw,Buckley:2009in}.

In this {\it Letter}, we examine the possible existence of a dark
force from a different perspective. Rather than limit its allowed
range of parameters based on observations, we show that it can {\em
ameliorate} tensions in astrophysical data. In particular, we find
that a Yukawa force in dark matter scattering would naturally produce
cores in dwarf galaxies while avoiding the myriad constraints on SIDM
which arise in systems with a much larger velocity dispersion, such as
clusters of galaxies.  The specific velocity dependence of the
interaction cross-section, as well as the possible exothermic nature
of the interaction, alleviate earlier concerns about the SIDM model.
To distinguish from previous approaches with a constant cross section
or a simple power law velocity dependence, we label this scenario as
Yukawa-Potential Interacting Dark Matter (YIDM).

\paragraph*{Dark Forces.}
The mediator of the force $\phi$ could be either a scalar or a vector,
as magnetic-type interactions are negligible.  The force could
couple to standard model fields through kinetic mixing with the
photon, or through mass mixing with the Higgs boson.  Constraints on
the presence of such a force come from a wide range of processes
\cite{Bjorken:2009mm,Schuster:2009au}, but ample parameter space
remains for a small mixing angle, $\epsilon \lsim 10^{-3}$. New
searches are underway to find precisely such a force carrier at $\sim
\gev$ energy experiments \cite{Essig:2010xa}.

Scattering through a massive mediator is equivalent to having a
Yukawa potential. The elastic scattering problem is then analogous to
the screened Coulomb scattering in a plasma \cite{PhysRevLett.90.225002},
which is well fit by a cross-section  \cite{Feng:2009hw,Finkbeiner:2010sm},
\begin{equation}
\vev{\sigma} \approx \left\{ \begin{array}{cc} \frac{4\pi}{m^2_\phi}
\beta^2 \ln(1+\beta^{-1}), & \beta\lsim 0.1, \\ \frac{8\pi}{m^2_\phi}
\beta^2/(1+1.5\beta^{1.65}), & 0.1 \lsim \beta \lsim 10^3, \\
\frac{\pi}{m^2_\phi} \left( \ln\beta + 1 - \frac{1}{2} \ln^{-1} \beta
\right)^2, & \beta \gsim 10^3,
\end{array} \right.
\label{eq:momtranssigma}
\end{equation}
where $\beta = \pi v_{\sigma}^2/v^2=2 \alpha_d m_\phi / (m_\chi v^2)$,
and $v$ is the relative velocity of the particles. We use angular
brackets to denote that this is the momentum-transfer weighted cross
section. Here, $v_{\sigma}$ is the velocity at which the
momentum-weighted scattering rate $\vev{\sigma v}$ peaks at a cross
section value of $\sigma_{\rm max} = 22.7/m_\phi^2$. The above
expression can be approximately generalized to the inelastic case by
substituting $m_\phi \rightarrow \sqrt{m_\chi \delta}$ for the
characteristic minimum momentum transfer when $m_\phi< \sqrt{m_\chi
\delta}$ (see discussion in \cite{Finkbeiner:2010sm}). This expression is derived using classical physics, and thus, it is important to note what quantum effects can come into play. In cases where the de Broglie wavelength is longer than the Compton wavelength of the force $m_\phi^{-1}$, the quantum calculation should be considered for quantitative results. Nonetheless, the same qualitative features should remain: the cross section should saturate at low velocities near $\sigma \sim m_\phi^{-2}$, and at high velocities, where the classical approximation is valid, it should fall rapidly.

  Figure 1 depicts the velocity dependence of the elastic cross-section in
Eq. (\ref{eq:momtranssigma}). Interestingly, the scattering rate is
nearly constant at low velocities, peaks at a velocity $v_\sigma$, and
declines sharply at $v>v_\sigma$, allowing it to introduce cores in
dwarf galaxies where the velocity dispersion is low ($v\sim 10~{\rm
km~s^{-1}}$) but not in clusters of galaxies where the characteristic
velocities are larger by two orders of magnitude ($v\sim 10^3~{\rm
km~s^{-1}}$). The normalizations of the cross-section and velocity are
determined by free parameters in the interaction Lagrangian (see
Appendix), with the Compton wavelength of the interaction setting the
relevant spatial scale.  We show two possible values of the peak
velocity, one that would produce cores only in dwarf galaxies
($v_\sigma=10~{\rm km~s^{-1}}$), and another that would produce cores
in more massive galaxies ($v_\sigma=10^2~{\rm km~s^{-1}}$) as implied
by data on low surface brightness galaxies \cite{Naray}.  At any given
halo mass, we expect scatter in the core properties of individual
halos, due to variations in their age and assembly history.

Having one collision per Hubble time at the characteristic core
density of dwarf galaxies $\sim 0.1M_\odot~{\rm pc}^{-3}$, translates
to the condition $({m_\chi}/{10 \gev})({m_\phi}/{100 \mev})^2 \sim1$
(see Appendix). An order of magnitude larger cross-sections are also
allowed by the data. Figure 2 shows the allowed parameter ranges
\cite{Buckley:2009in} that would naturally explain the dark matter
distribution in observed astrophysical objects.  We find that even
though collisions shape the central profiles of dwarf galaxies, the
standard collisionless treatment still provides an excellent
approximation for the dark matter dynamics in X-ray clusters.

\begin{figure}[t]
\centerline{\includegraphics[width=0.45 \textwidth]{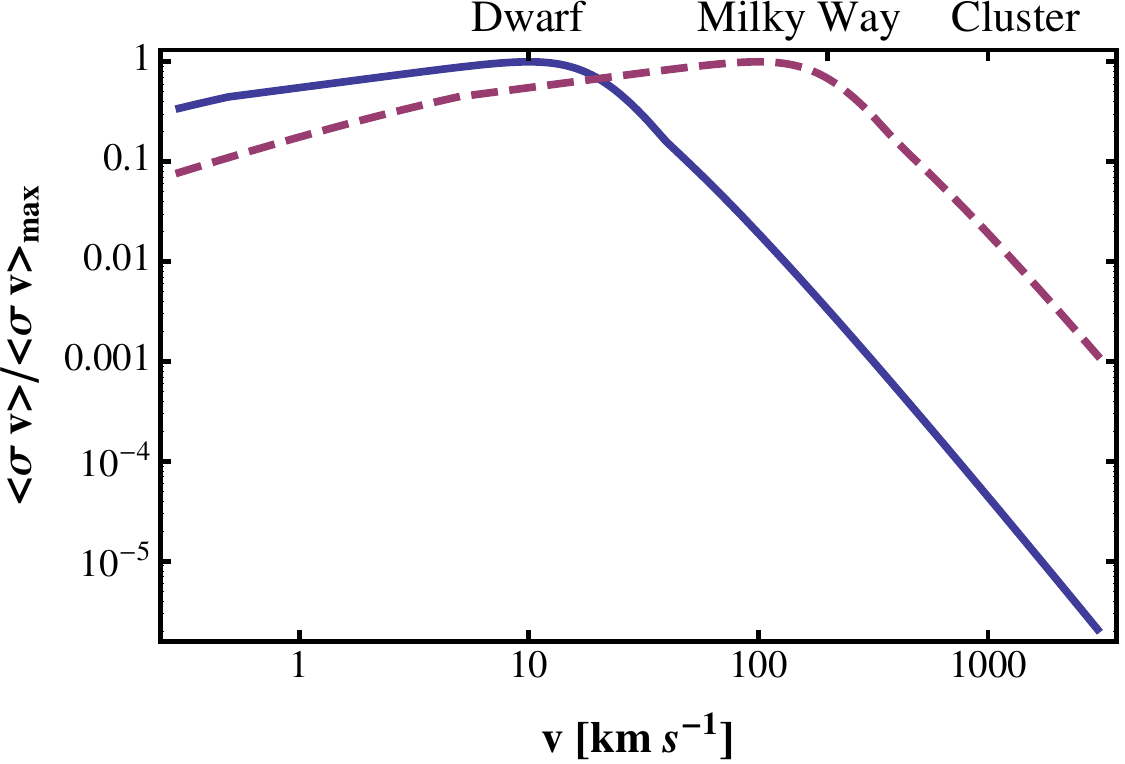}\hskip 0.2in}
\caption{Dependence of the self-interaction cross-section ($\sigma$)
on the relative velocity ($v$) for dark matter interacting through a
Yukawa potential. The normalizations of $\sigma$ and $v$ are set by
free parameters in the underlying Lagrangian (see Appendix), and we
show two possible curves peaking at $v_\sigma = 10~{\rm km~s^{-1}}$
and $=100~{\rm km~s^{-1}}$ ({\em blue, solid} and {\em purple,
dashed}, respectively). }
\end{figure}

\begin{figure}[th]
\centerline{\includegraphics[width=0.45 \textwidth]{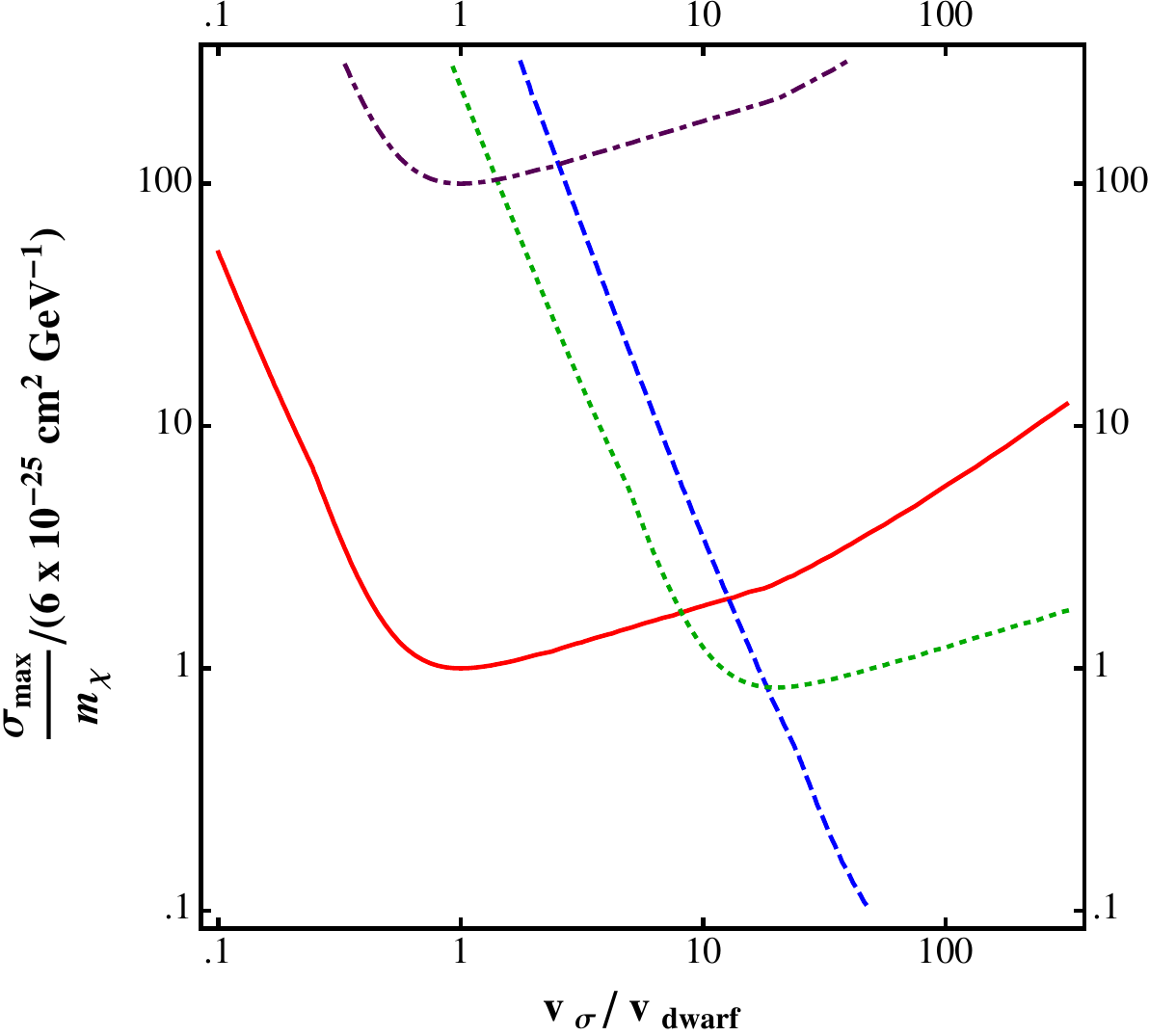}}
\caption{Astrophysical constraints on the normalization of the
self-interaction cross-section ($\sigma_{\rm max}$) as a function of
the velocity at which the peak collision rate is obtained
($v_{\sigma}$) in Fig. 1. The {\em red solid} line is normalized to
have $\vev{\sigma}_{max}/m_\chi \approx 6\times 10^{-25}~{\rm
cm^2/GeV}$ at $v_{\rm dwarf}\approx 10~{\rm km~s^{-1}}$, which should
be regarded as the minimum interaction necessary to flatten the cores
of dwarf galaxies.  Additional lines indicate upper limits on the
cross-section based on astrophysical considerations: X-ray cluster
ellipticity ({\em blue, dashed}), limiting $(\sigma_{max}/m_\chi)
\lsim 4\times 10^{-26}~{\rm cm^2/GeV}$ at $v\sim 10^3 {\rm
km~s^{-1}}$; destruction of dwarf sub-halos through collisions with
high velocity particles from a larger parent halo in which these
dwarfs are embedded ({\em green, dotted}), limiting
$(\sigma_{max}/m_\chi)\lsim 5 \times 10^{-25}~{\rm cm^2/GeV}$
\cite{Gnedin:2000ea} at $v\sim 200 {\rm km~s^{-1}}$; and requiring the
number of scatters in dwarfs to be less than $\sim 10^2$ during the
age of the Universe to avoid the gravothermal catastrophe ({\em
purple, dash-dotted}). Related limits are summarized in
\cite{Buckley:2009in}.}
\end{figure}

\paragraph*{Exothermic interactions.}
The presence of excited states related to a new force has important
implications for the properties of dark matter. In particular,
upscattering (``inelastic dark matter'' (iDM)
\cite{TuckerSmith:2001hy}) or downscattering
\cite{Finkbeiner:2009mi,Batell:2009vb,Lang:2010cd,Essig:2010ye,Graham:2010ca}
off nuclei can dramatically affect direct detection experiments. Dark
matter self scattering into an excited state (``eXciting Dark Matter''
(XDM) \cite{Finkbeiner:2007kk}) has been invoked to explain the excess
511 keV flux observed by INTEGRAL
\cite{Finkbeiner:2007kk,Pospelov:2007xh,Chen:2009dm,Finkbeiner:2009mi,Chen:2009av,Chen:2009dm}.

We focus here on the response of dwarf galaxies to the presence of
excited states, YIDM$^*$, which we assume are copiously present from
the early universe.  In particular, the release of kinetic energy in
YIDM$^*$ collisions would help to evade the gravothermal catastrophe
\cite{Balberg}, in just the same way that the energy released by
primordial stellar binaries weakens core collapse in globular clusters
\cite{Goodman}.

If excited states exist, then a major fraction of the CDM might be
excited when the WIMPS decouple thermally in the early Universe. This
excitation could be stable on cosmological times in models where the
dark force mixes with electromagnetism
\cite{Finkbeiner:2009mi,Batell:2009vb}. The scattering process can
decouple in the early universe at temperatures above the splitting,
leaving essentially equal abundances of all states of the dark
matter. Alternatively, the excited states could be produced
non-thermally. The former scenario tends to require light
($m_\chi\sim$ MeV) particles, while the latter would be more naturally
weak-scale.  After decoupling, the early CDM dynamics is identical to
the standard collisionless model until dwarf galaxies form and the
crossing of dark matter streams at low $v$ occurs in their cores,
giving rise to self-interactions on a timescale shorter than the age
of the Universe.

We denote the velocity imparted to an initially slowly-moving WIMPs
($v<v_{\rm crit}$) upon scattering by $v_{\rm
crit}=\sqrt{\delta/m}$. Since the gravitational potential of a dwarf
galaxy halo is shallow, sufficiently exothermic collisions could eject
colliding particles out of the halo if $v_{\rm crit}$ exceeds the
local escape velocity. If $\chi \chi^* \rightarrow \chi \chi$
scatterings can occur (see Appendix), then the halo core will lose
particles until it reaches a density such that the interaction time is
comparable to the age of the halo. Requiring that the final core
particles will interact only once over the current age of the Universe
yields a final mass density of dark matter, \be \rho_\chi\sim 0.1
{M_\odot \over {\rm pc}^{3}}\left({\sigma /6\! \times\!  10^{-25} {\rm
cm^2}\over {m_\chi/{\rm GeV}}}\right)^{-1}\left(\frac{v}{10~{\rm
km~s^{-1}}}\right)^{-1} .  \ee The profile of $\rho_\chi$ will then be
set by the velocity dependence of $\vev{\sigma v}$ and the
gravitational potential in a steady state. Below we show that for an
exothermic interaction $\vev{\sigma v}$ is constant, leading naturally
to a constant density core in dwarf galaxies.  This model predicts
that dwarf galaxies of a similar age should have a similar core
density, in agreement with interpretations of data for the nearest
dwarf galaxies \cite{Strigari}.

If the characteristic scattering velocity is much higher than $v_{\rm
crit}$, the process is essentially elastic and one can employ
Eq. (\ref{eq:momtranssigma}). On the other hand, if the process is
highly exothermic then the scattering rate is given simply by $\sigma
v = 2 \pi v_{\rm crit} \alpha_d^2/m_\phi^4$ for $m_\chi \delta <
m_\phi^2$ (assuming the scattering process is still perturbative,
i.e. $\beta<1$).  The resulting velocity-independent $\sigma v$ would
naturally produce cores with a flat density profile in dwarf galaxies.

The density flattening in dwarf galaxies does not imply the same upper
limit on the dark matter density in all its cusps.  For massive halos,
the release of excess kinetic energy by collisions has a marginal
significance, since it only perturbs the low-velocity tail of the CDM
distribution function and adds a negligible energy at high relative
velocities where the majority of particles have a low interaction rate
anyway.

The evaporation of exothermic YIDM$^*$ from dwarf galaxies with a
gravitational binding energy below the energy released in collisions
could potentially account for the deficit in the observed abundance of
dwarf galaxies relative to theoretical CDM expectations
\cite{Bullock}.  Numerical simulations are necessary to reliably
quantify this important effect. Also, dark matter halos which accrete
cold gas at early times but evaporate at late time might leave behind
a star cluster with little dark matter. If so, some old globular clusters
\cite{Conroy} might be the sought-after remnants of the missing dwarf
galaxies in the Milky Way halo.

\paragraph*{Evolution with redshift.} 
The primordial density perturbations are modified by WIMP scatterings,
but for $m_\chi \gsim \gev$ this modification ends well before
observable modes enter the horizon, leaving the standard nearly
scale-invariant power-spectrum of density perturbations.  The imprint
of collisions on the density profile of halos is expected to evolve
with redshift, because at earlier cosmic times halos are denser and
younger.
A halo of mass $M$ collapsing at a redshift $z\gg 1$ has a
characteristic virial radius \cite{Book}, \beq r_{\rm vir}=1.5
\left(\frac{M}{10^8 M_{\sun} }\right)^{1/3} \left (\frac{1+z}{10}
\right)^{-1} \ {\rm kpc}\ ,
\label{rvir}\eeq a corresponding circular velocity, \beq
V_{\rm vir}=17.0 \left( \frac{M}{10^8 M_{\sun} }\right)^{1/3} \left(
\frac{1+z} {10} \right)^{1/2}\ {\rm km~s}^{-1}\ , \label{Vceqn} \eeq
and an age limit of $\sim 0.5~{\rm Gyr}[(1+z)/10]^{-3/2}$. 
It would be particularly interesting to explore the formation of the
first galaxies using numerical simulations of the YIDM
model. Deviations from the standard CDM predictions could be tested by
upcoming galaxy surveys or 21-cm observations of the high-redshift
Universe \cite{Book,Pritchard}. For example, the scaling $\sigma
\propto v^{-4}$ might not be allowed to continue to arbitrarily low
velocities, as this would have delayed reionization beyond
observational constraints \cite{2010MNRAS.408...57P}.

\paragraph*{Discussion.}
We have shown that a Yukawa potential interaction of dark matter can
explain the recent data on dark matter cores in dwarf galaxies
\cite{Oh,Walker}, while evading the many constraints previously
considered for SIDM. The new ingredients of the YIDM model involve the
non-trivial velocity dependence of the scattering rate and
the possibility of exothermic interactions.

The velocity dependence of the cross-section does not have a simple
power-law form, as invoked previously \cite{Yoshida,Gnedin:2000ea}.
The presence of a plateau with a sharp cut-off in the scattering rate
of YIDM allows the interaction to be effective for dwarf galaxies
while being entirely suppressed at high velocities relevant for
cluster cores and the evaporation of small sub-halos within bigger
halos.  At the same time, it does not rise indefinitely at low
velocities and thus avoid other concerns \cite{Gnedin:2000ea}.

Excited states naturally accompany a dark force and introduce a
qualitatively new ingredient: the possibility of energy release. In
deep gravitational potentials the scatterings will be similar to
the standard elastic case, whereas in shallow gravitational potentials a
single scatter can eject the colliding dark matter particles from the
halo. Exothermic interactions drive dwarf halos to constant density
cores. Moreover, for the earliest forming halos with the highest dark
matter density, it is possible that the dark matter would evaporate at
later times, leaving behind baryonically-dominated stellar clusters
such as the oldest globular clusters.

Unlike SIDM with a single parameter of $\sigma/m_\chi$, YIDM has a
broader parameter space. Out of the possible combinations of the
underlying particle physics parameters ($\alpha, \delta, m_\chi,
m_\phi$), only three affect the properties of halos: the cross-section
$\sigma_{\rm max}$ and velocity $v_\sigma$ at which the scattering
rate $\vev{\sigma v}$ peaks; and in the ${\rm YIDM^*}$ scenario there
is the additional parameter of the velocity imparted to a particle
undergoing an exothermic scattering, $v_{\rm crit}$.

Since halos form hierarchically and the coarse-grained phase-space
density can only decrease through their mergers (Liouville's theorem),
the development of a core in dwarf galaxies would trim the central
cusp in bigger halos as well. Numerical simulations are required to
quantify the average behavior as well as the scatter in the core
properties of halos as a function of mass and redshift.

Finally, we note that the similarity between the required
self-interaction cross-section per unit mass of the dark matter and
baryons may indicate a deep underlying relationship between these
components.  

\begin{appendix}
\section{Appendix: Particle Physics Implications}

Although a thorough investigation of the particle physics parameter
space is beyond our scope here, it is worthwhile to at least briefly
consider the range of parameter space available to us.  As mentioned,
the mediator of the force $\phi$ could be either a scalar or a vector,
since magnetic-type interactions are negligible.  Simple perturbative
realizations of dark forces involve models with an additional boson
$\phi_\mu$ of a small mass $m_\phi \lsim \rm GeV$.  We can assume that
the WIMPs are charged under a new $U(1)_{\rm dark}$ symmetry and allow
the possibility of a small splitting between their Majorana states,
leading to the Lagrangian, \ba {\cal L} = \bar \chi \not \! \! D \chi
& +& \frac{1}{4}F^d_{\mu\nu} F^{d\mu\nu}+\epsilon F_{\mu\nu}
F^{d\mu\nu}+m_\phi^2 \phi_{ \mu} \phi^{\mu}\nonumber \\ &+ & m_\chi^2
\bar \chi \chi + \delta \chi \chi .
\label{eq:lagrangian}
\aa This can be trivially generalized to a non-Abelian model with
multiple excited states \cite{ArkaniHamed:2008qn,Baumgart:2009tn},
which induces splittings between the states radiatively at order
$\alpha_d m_\phi$, where $\alpha_d$ is the fine structure constant of
the dark force.  Similarly, the force can arise through a scalar
exchange with the additional terms \ba {\cal L} \supset y \bar \chi
\chi \phi & +& \partial_{\mu} \phi \partial^\mu \phi^* + m_\phi^2
\phi^* \phi.
\label{eq:scalarlagrangian}
\aa Here, the effective coupling constant $\alpha_d = y^2/4
\pi$. Elastic scatterings can arise for both scalar and
vector, while inelastic scatterings are most natural in the gauged
case.  We begin by focusing on the elastic scattering scenario.

The natural mass scale that determines the maximum momentum transfer
cross section is $m_\phi$. For large values of $0.1\lsim \beta\lsim
10^3$ (above which the cross section in galaxy clusters is $\gsim
10^{-3}$ that in dwarf galaxies), Eq. (\ref{eq:momtranssigma}) yields
$\langle \sigma \rangle \approx (16 \pi /m_\phi^2) \beta^2/(2 + 3
\beta^{1.65})$. Requiring a cross section/mass of $6\times 10^{-25}{\rm
cm^2~GeV^{-1}}$ then implies $m_\chi m_\phi^2 = 3\times 10^{-2}
\beta^2/(2+3 \beta^{1.65})~{\rm GeV^{3}}$. Thus, for $m_\chi = 10 \,
(10^3) \, \gev$, one has $m_\phi < 10^2 \, (10)\, \mev$. For standard
WIMPs interacting through weak scale forces, such phenomena do not
occur; however, for dark matter augmented by light dark forces it is
possible. For lighter ($\sim \gev$) WIMPs, a quantum calculation should 
be used to generate the precise quantitative results. The relevant mass range for $\phi$ is not surprising in light of the relatively
weak constraints from galactic dynamics found by recent studies
\cite{Feng:2009hw,Buckley:2009in}.

Adopting $\beta= \beta_{\rm dwarf}$ in a typical dwarf galaxy implies
through the above relation, $\alpha_d \approx 10^{-7}(m_\chi/{\rm
GeV})^{3/2} \sqrt{3 \beta_{\rm dwarf}^{1.65}/2+1}$. The dependence on
$\beta_{\rm dwarf}$ and $m_\chi$ allows a wide range of values,
$10^{-5} \lsim \alpha_d \lsim 1$, with the low end for $\beta_{\rm
dwarf}\sim 1$ and $m_\chi \sim 10\gev$, and the high end for
$\beta_{\rm dwarf} \sim 10^3$ and $m_\chi \sim 10^3 \gev$.

Although $\alpha_d$ can in principle have any value in the above
range, experience in the standard model suggests that gauge fields
have coupling constants $\alpha_d \gsim 10^{-3}$, while forces arising
from scalar exchange could have a much wider range (down to $y^2/4\pi
\sim 10^{-12}$). Thus, it is instructive to consider whether we have
any additional constraints from cosmology. One possible assumption is
that the dark matter is a thermal relic. This is a strong assumption,
with compelling alternative models of asymmetric dark matter that
relate the dark matter density to the baryon asymmetry (see,
e.g. Refs. \cite{Kaplan:1991ah,Kitano:2004sv,Farrar:2005zd,Kaplan:2009ag,Chang:2009sv,Cohen:2010kn}),
but it is nevertheless a general category worth examining.

The basic requirement of a thermal relic is that the average
annihilation cross section $\vev{\sigma_{\rm ann} v} \approx 3 \times
10^{-26}{\rm cm^3~s^{-1}}$ at decoupling. Annihilation into new bosons
is characterized by a rate constant $\vev{\sigma_{\rm ann}v} \approx {\pi
\alpha^2}/{m_\chi^2}$, but it is important to recognize that there can
be many additional annihilation channels into other charged states or
even into other force carriers that simply happen not to be the
dominant contributors to present-day scattering. Thus, this should be
properly taken as a upper bound on $\alpha_d$. Thus, all but the
largest values of $\alpha_d$ are acceptable from the perspective of
thermal freezeout, which would require a primordial asymmetry.

It is important to emphasize that for our purposes, it is the lightest
mediator that will set the characteristic scattering scale, not
necessarily the one with the largest coupling constant, simply because
of the saturation that sets in at low velocities. Thus, it could well
be that the underlying Lagrangian is described by a gauge theory,
spontaneously broken by a Higgs field, and the light Higgs through a
small Yukawa $\phi$ could mediate the self-interaction, even if the
gauge interaction is mediated by a field too heavy to induce CDM
scatterings.

For the exothermic scenario, the parameter range for dark matter mass
is much narrower. We can determine this range by requiring that the
scattering in the early universe had ceased by the time the CDM had
reached a low enough temperature to thermally deplete the excited
states through the same size cross section that is relevant today,
i.e. \be \rho_\chi \frac{\langle \sigma v \rangle}{m_\chi} \lsim 3 H, \ee after the
time at which $v_\chi< v_{\rm dwarf}\approx 10~{\rm km\, s^{-1}}$
(where $H$ is the Hubble parameter). The temperature $T_\chi$ at which
the WIMPs slow down to the speed in current dwarf galaxies is
determined by the temperature $T_{\rm dec}$ at which they decouple
from the photon bath, $T_\chi \sim T_{\gamma}^2/T_{\rm dec}$, where
$T_{\gamma}\sim T_0(1+z)$. Thus, setting $\rho_\chi = \Omega_\chi \rho_{c0}
(T_\gamma/T_0)^{3}$, where $\rho_{c0}$ is the present-day critical density
and $H^2 \sim (8\pi G/3) g_* T_\gamma^4$, yields
$m_\chi \lsim m_e \times {\rm MeV}/ {T_{\rm dec}}$. Since it would
be unnatural to have $T_{\rm dec}\ll m_e$ (as we assume the
WIMPs are not charged), this suggests WIMPs in the sub-MeV mass
range. In such a case, it would be most natural to have an asymmetric
model such as described in Ref. \cite{Cohen:2010kn}.

Another possibility would be that the excited states are
produced non-thermally, i.e., a heavier state freezes out and
populates the excited states after they are adequately dilute that they
cannot down-scatter on themselves until the present day. In this
case, our expectation would be for characteristically weak-scale
masses in order to have the appropriate relic abundance.

From a model building perspective, such scenarios would most naturally
be constructed within non-Abelian gauge theories. This is because
for Abelian gauge theories, the scattering always changes the number
of excited states by a multiple of two, i.e., $11\Longleftrightarrow
22$ and $12\Longleftrightarrow 21$. This suggests that it will be
difficult to inelastically scatter the $1$ states.  However, in the
case of a non-Abelian theory, more states, and thus more scattering
possibilities, are present. For instance, $13\Longrightarrow 22$ would
allow an exothermic scattering for even ground-state particles.

In summary, there is a significant but nonetheless constrained range
of parameters which could yield the YIDM and ${\rm YIDM^*}$
phenomenology. We leave detailed model building for future work.

\end{appendix}

\bigskip
\bigskip
\paragraph*{Acknowledgments.}
We thank Doug Finkbeiner, Patrick Fox and Matt Walker for helpful comments on the manuscript.  This
work was supported in part by NSF grant AST-0907890 and NASA grants
NNX08AL43G and NNA09DB30A for AL. NW is supported by DOE OJI grant
\#DE-FG02-06ER41417 and NSF grant \#0947827, as well as by the Amborse
Monell Foundation.

\bibliography{cores}

\begin{thebibliography}{56}
\expandafter\ifx\csname natexlab\endcsname\relax\def\natexlab#1{#1}\fi
\expandafter\ifx\csname bibnamefont\endcsname\relax
  \def\bibnamefont#1{#1}\fi
\expandafter\ifx\csname bibfnamefont\endcsname\relax
  \def\bibfnamefont#1{#1}\fi
\expandafter\ifx\csname citenamefont\endcsname\relax
  \def\citenamefont#1{#1}\fi
\expandafter\ifx\csname url\endcsname\relax
  \def\url#1{\texttt{#1}}\fi
\expandafter\ifx\csname urlprefix\endcsname\relax\def\urlprefix{URL }\fi
\providecommand{\bibinfo}[2]{#2}
\providecommand{\eprint}[2][]{\url{#2}}

\bibitem[{\citenamefont{Komatsu et~al.}(2010)}]{WMAP}
\bibinfo{author}{\bibfnamefont{E.}~\bibnamefont{Komatsu}} \bibnamefont{et~al.}
  (\bibinfo{year}{2010}), \eprint{1001.4538}.

\bibitem[{\citenamefont{{de Blok}}(2010)}]{Blok}
\bibinfo{author}{\bibfnamefont{W.~J.~G.} \bibnamefont{{de Blok}}},
  \bibinfo{journal}{Advances in Astronomy} \textbf{\bibinfo{volume}{2010}}
  (\bibinfo{year}{2010}), \eprint{0910.3538}.

\bibitem[{\citenamefont{{Navarro} et~al.}(2010)\citenamefont{{Navarro},
  {Ludlow}, {Springel}, {Wang}, {Vogelsberger}, {White}, {Jenkins}, {Frenk},
  and {Helmi}}}]{NFW}
\bibinfo{author}{\bibfnamefont{J.~F.} \bibnamefont{{Navarro}}},
  \bibinfo{author}{\bibfnamefont{A.}~\bibnamefont{{Ludlow}}},
  \bibinfo{author}{\bibfnamefont{V.}~\bibnamefont{{Springel}}},
  \bibinfo{author}{\bibfnamefont{J.}~\bibnamefont{{Wang}}},
  \bibinfo{author}{\bibfnamefont{M.}~\bibnamefont{{Vogelsberger}}},
  \bibinfo{author}{\bibfnamefont{S.~D.~M.} \bibnamefont{{White}}},
  \bibinfo{author}{\bibfnamefont{A.}~\bibnamefont{{Jenkins}}},
  \bibinfo{author}{\bibfnamefont{C.~S.} \bibnamefont{{Frenk}}},
  \bibnamefont{and} \bibinfo{author}{\bibfnamefont{A.}~\bibnamefont{{Helmi}}},
  \bibinfo{journal}{\mnras} \textbf{\bibinfo{volume}{402}}, \bibinfo{pages}{21}
  (\bibinfo{year}{2010}), \eprint{0810.1522}.

\bibitem[{\citenamefont{Oh et~al.}(2010{\natexlab{a}})\citenamefont{Oh,
  de~Blok, Brinks, Walter, and Kennicutt}}]{Oh}
\bibinfo{author}{\bibfnamefont{S.-H.} \bibnamefont{Oh}},
  \bibinfo{author}{\bibfnamefont{W.~J.~G.} \bibnamefont{de~Blok}},
  \bibinfo{author}{\bibfnamefont{E.}~\bibnamefont{Brinks}},
  \bibinfo{author}{\bibfnamefont{F.}~\bibnamefont{Walter}}, \bibnamefont{and}
  \bibinfo{author}{\bibfnamefont{J.}~\bibnamefont{Kennicutt},
  \bibfnamefont{Robert~C.}} (\bibinfo{year}{2010}{\natexlab{a}}),
  \eprint{1011.0899}.

\bibitem[{\citenamefont{Goerdt et~al.}(2006)\citenamefont{Goerdt, Moore, Read,
  Stadel, and Zemp}}]{Goerdt}
\bibinfo{author}{\bibfnamefont{T.}~\bibnamefont{Goerdt}},
  \bibinfo{author}{\bibfnamefont{B.}~\bibnamefont{Moore}},
  \bibinfo{author}{\bibfnamefont{J.~I.} \bibnamefont{Read}},
  \bibinfo{author}{\bibfnamefont{J.}~\bibnamefont{Stadel}}, \bibnamefont{and}
  \bibinfo{author}{\bibfnamefont{M.}~\bibnamefont{Zemp}},
  \bibinfo{journal}{Mon. Not. Roy. Astron. Soc.}
  \textbf{\bibinfo{volume}{368}}, \bibinfo{pages}{1073} (\bibinfo{year}{2006}),
  \eprint{astro-ph/0601404}.

\bibitem[{\citenamefont{Kleyna et~al.}(2003)\citenamefont{Kleyna, Wilkinson,
  Gilmore, and Evans}}]{Kleyna}
\bibinfo{author}{\bibfnamefont{J.~T.} \bibnamefont{Kleyna}},
  \bibinfo{author}{\bibfnamefont{M.~I.} \bibnamefont{Wilkinson}},
  \bibinfo{author}{\bibfnamefont{G.}~\bibnamefont{Gilmore}}, \bibnamefont{and}
  \bibinfo{author}{\bibfnamefont{N.~W.} \bibnamefont{Evans}},
  \bibinfo{journal}{Astrophys. J.} \textbf{\bibinfo{volume}{588}},
  \bibinfo{pages}{L21} (\bibinfo{year}{2003}), \eprint{astro-ph/0304093}.

\bibitem[{\citenamefont{Walker and Penarrubia}(2010)}]{Walker}
\bibinfo{author}{\bibfnamefont{M.}~\bibnamefont{Walker}} \bibnamefont{and}
  \bibinfo{author}{\bibfnamefont{J.}~\bibnamefont{Penarrubia}},
  \bibinfo{journal}{submitted for publication}  (\bibinfo{year}{2010}).

\bibitem[{\citenamefont{Gnedin and Zhao}(2002)}]{Gnedin}
\bibinfo{author}{\bibfnamefont{O.~Y.} \bibnamefont{Gnedin}} \bibnamefont{and}
  \bibinfo{author}{\bibfnamefont{H.}~\bibnamefont{Zhao}},
  \bibinfo{journal}{Mon. Not. Roy. Astron. Soc.}
  \textbf{\bibinfo{volume}{333}}, \bibinfo{pages}{299} (\bibinfo{year}{2002}),
  \eprint{astro-ph/0108108}.

\bibitem[{\citenamefont{Governato et~al.}(2009)}]{Gov}
\bibinfo{author}{\bibfnamefont{F.}~\bibnamefont{Governato}}
  \bibnamefont{et~al.} (\bibinfo{year}{2009}), \eprint{0911.2237}.

\bibitem[{\citenamefont{Oh et~al.}(2010{\natexlab{b}})}]{Oh:2010mc}
\bibinfo{author}{\bibfnamefont{S.-H.} \bibnamefont{Oh}} \bibnamefont{et~al.}
  (\bibinfo{year}{2010}{\natexlab{b}}), \eprint{1011.2777}.

\bibitem[{\citenamefont{Blumenthal et~al.}(1986)\citenamefont{Blumenthal,
  Faber, Flores, and Primack}}]{Blum}
\bibinfo{author}{\bibfnamefont{G.~R.} \bibnamefont{Blumenthal}},
  \bibinfo{author}{\bibfnamefont{S.~M.} \bibnamefont{Faber}},
  \bibinfo{author}{\bibfnamefont{R.}~\bibnamefont{Flores}}, \bibnamefont{and}
  \bibinfo{author}{\bibfnamefont{J.~R.} \bibnamefont{Primack}},
  \bibinfo{journal}{Astrophys. J.} \textbf{\bibinfo{volume}{301}},
  \bibinfo{pages}{27} (\bibinfo{year}{1986}).

\bibitem[{\citenamefont{Sawala et~al.}(2010)\citenamefont{Sawala, Guo,
  Scannapieco, Jenkins, and White}}]{White}
\bibinfo{author}{\bibfnamefont{T.}~\bibnamefont{Sawala}},
  \bibinfo{author}{\bibfnamefont{Q.}~\bibnamefont{Guo}},
  \bibinfo{author}{\bibfnamefont{C.}~\bibnamefont{Scannapieco}},
  \bibinfo{author}{\bibfnamefont{A.}~\bibnamefont{Jenkins}}, \bibnamefont{and}
  \bibinfo{author}{\bibfnamefont{S.~D.~M.} \bibnamefont{White}}
  (\bibinfo{year}{2010}), \eprint{1003.0671}.

\bibitem[{\citenamefont{{Guo} et~al.}(2010)\citenamefont{{Guo}, {White},
  {Boylan-Kolchin}, {De Lucia}, {Kauffmann}, {Lemson}, {Li}, {Springel}, and
  {Weinmann}}}]{Guo}
\bibinfo{author}{\bibfnamefont{Q.}~\bibnamefont{{Guo}}},
  \bibinfo{author}{\bibfnamefont{S.}~\bibnamefont{{White}}},
  \bibinfo{author}{\bibfnamefont{M.}~\bibnamefont{{Boylan-Kolchin}}},
  \bibinfo{author}{\bibfnamefont{G.}~\bibnamefont{{De Lucia}}},
  \bibinfo{author}{\bibfnamefont{G.}~\bibnamefont{{Kauffmann}}},
  \bibinfo{author}{\bibfnamefont{G.}~\bibnamefont{{Lemson}}},
  \bibinfo{author}{\bibfnamefont{C.}~\bibnamefont{{Li}}},
  \bibinfo{author}{\bibfnamefont{V.}~\bibnamefont{{Springel}}},
  \bibnamefont{and}
  \bibinfo{author}{\bibfnamefont{S.}~\bibnamefont{{Weinmann}}},
  \bibinfo{journal}{ArXiv e-prints}  (\bibinfo{year}{2010}),
  \eprint{1006.0106}.

\bibitem[{\citenamefont{{Sawala} et~al.}(2010)\citenamefont{{Sawala},
  {Scannapieco}, {Maio}, and {White}}}]{2010MNRAS.402.1599S}
\bibinfo{author}{\bibfnamefont{T.}~\bibnamefont{{Sawala}}},
  \bibinfo{author}{\bibfnamefont{C.}~\bibnamefont{{Scannapieco}}},
  \bibinfo{author}{\bibfnamefont{U.}~\bibnamefont{{Maio}}}, \bibnamefont{and}
  \bibinfo{author}{\bibfnamefont{S.}~\bibnamefont{{White}}},
  \bibinfo{journal}{\mnras} \textbf{\bibinfo{volume}{402}},
  \bibinfo{pages}{1599} (\bibinfo{year}{2010}), \eprint{0902.1754}.

\bibitem[{\citenamefont{Spergel and Steinhardt}(2000)}]{SIDM}
\bibinfo{author}{\bibfnamefont{D.~N.} \bibnamefont{Spergel}} \bibnamefont{and}
  \bibinfo{author}{\bibfnamefont{P.~J.} \bibnamefont{Steinhardt}},
  \bibinfo{journal}{Phys. Rev. Lett.} \textbf{\bibinfo{volume}{84}},
  \bibinfo{pages}{3760} (\bibinfo{year}{2000}), \eprint{astro-ph/9909386}.

\bibitem[{\citenamefont{{Carlson} et~al.}(1992)\citenamefont{{Carlson},
  {Machacek}, and {Hall}}}]{Carlson}
\bibinfo{author}{\bibfnamefont{E.~D.} \bibnamefont{{Carlson}}},
  \bibinfo{author}{\bibfnamefont{M.~E.} \bibnamefont{{Machacek}}},
  \bibnamefont{and} \bibinfo{author}{\bibfnamefont{L.~J.}
  \bibnamefont{{Hall}}}, \bibinfo{journal}{\apj}
  \textbf{\bibinfo{volume}{398}}, \bibinfo{pages}{43} (\bibinfo{year}{1992}).

\bibitem[{\citenamefont{{Machacek} et~al.}(1993)\citenamefont{{Machacek},
  {Carlson}, and {Hall}}}]{machecek}
\bibinfo{author}{\bibfnamefont{M.~E.} \bibnamefont{{Machacek}}},
  \bibinfo{author}{\bibfnamefont{E.~D.} \bibnamefont{{Carlson}}},
  \bibnamefont{and} \bibinfo{author}{\bibfnamefont{L.~J.}
  \bibnamefont{{Hall}}}, in \emph{\bibinfo{booktitle}{Texas/PASCOS '92:
  Relativistic Astrophysics and Particle Cosmology}}, edited by
  \bibinfo{editor}{\bibnamefont{{C.~W.~Akerlof \& M.~A.~Srednicki}}}
  (\bibinfo{year}{1993}), vol. \bibinfo{volume}{688} of
  \emph{\bibinfo{series}{Annals of the New York Academy of Sciences}}, pp.
  \bibinfo{pages}{681--+}.

\bibitem[{\citenamefont{{de Laix} et~al.}(1995)\citenamefont{{de Laix},
  {Scherrer}, and {Schaefer}}}]{laix}
\bibinfo{author}{\bibfnamefont{A.~A.} \bibnamefont{{de Laix}}},
  \bibinfo{author}{\bibfnamefont{R.~J.} \bibnamefont{{Scherrer}}},
  \bibnamefont{and} \bibinfo{author}{\bibfnamefont{R.~K.}
  \bibnamefont{{Schaefer}}}, \bibinfo{journal}{\apj}
  \textbf{\bibinfo{volume}{452}}, \bibinfo{pages}{495} (\bibinfo{year}{1995}),
  \eprint{arXiv:astro-ph/9502087}.

\bibitem[{\citenamefont{{Miralda-Escud{\'e}}}(2002)}]{Miralda}
\bibinfo{author}{\bibfnamefont{J.}~\bibnamefont{{Miralda-Escud{\'e}}}},
  \bibinfo{journal}{\apj} \textbf{\bibinfo{volume}{564}}, \bibinfo{pages}{60}
  (\bibinfo{year}{2002}).

\bibitem[{\citenamefont{Yoshida et~al.}(2000)\citenamefont{Yoshida, Springel,
  White, and Tormen}}]{Yoshida}
\bibinfo{author}{\bibfnamefont{N.}~\bibnamefont{Yoshida}},
  \bibinfo{author}{\bibfnamefont{V.}~\bibnamefont{Springel}},
  \bibinfo{author}{\bibfnamefont{S.~D.~M.} \bibnamefont{White}},
  \bibnamefont{and} \bibinfo{author}{\bibfnamefont{G.}~\bibnamefont{Tormen}},
  \bibinfo{journal}{Astrophys. J.} \textbf{\bibinfo{volume}{535}},
  \bibinfo{pages}{L103} (\bibinfo{year}{2000}), \eprint{astro-ph/0002362}.

\bibitem[{\citenamefont{Balberg et~al.}(2002)\citenamefont{Balberg, Shapiro,
  and Inagaki}}]{Balberg}
\bibinfo{author}{\bibfnamefont{S.}~\bibnamefont{Balberg}},
  \bibinfo{author}{\bibfnamefont{S.~L.} \bibnamefont{Shapiro}},
  \bibnamefont{and} \bibinfo{author}{\bibfnamefont{S.}~\bibnamefont{Inagaki}},
  \bibinfo{journal}{Astrophys. J.} \textbf{\bibinfo{volume}{568}},
  \bibinfo{pages}{475} (\bibinfo{year}{2002}), \eprint{astro-ph/0110561}.

\bibitem[{\citenamefont{Arkani-Hamed et~al.}(2009)\citenamefont{Arkani-Hamed,
  Finkbeiner, Slatyer, and Weiner}}]{ArkaniHamed:2008qn}
\bibinfo{author}{\bibfnamefont{N.}~\bibnamefont{Arkani-Hamed}},
  \bibinfo{author}{\bibfnamefont{D.~P.} \bibnamefont{Finkbeiner}},
  \bibinfo{author}{\bibfnamefont{T.~R.} \bibnamefont{Slatyer}},
  \bibnamefont{and} \bibinfo{author}{\bibfnamefont{N.}~\bibnamefont{Weiner}},
  \bibinfo{journal}{Phys. Rev.} \textbf{\bibinfo{volume}{D79}},
  \bibinfo{pages}{015014} (\bibinfo{year}{2009}), \eprint{0810.0713}.

\bibitem[{\citenamefont{Hogan and Dalcanton}(2000)}]{Hogan:2000bv}
\bibinfo{author}{\bibfnamefont{C.~J.} \bibnamefont{Hogan}} \bibnamefont{and}
  \bibinfo{author}{\bibfnamefont{J.~J.} \bibnamefont{Dalcanton}},
  \bibinfo{journal}{Phys. Rev.} \textbf{\bibinfo{volume}{D62}},
  \bibinfo{pages}{063511} (\bibinfo{year}{2000}), \eprint{astro-ph/0002330}.

\bibitem[{\citenamefont{Feng et~al.}(2010)\citenamefont{Feng, Kaplinghat, and
  Yu}}]{Feng:2009hw}
\bibinfo{author}{\bibfnamefont{J.~L.} \bibnamefont{Feng}},
  \bibinfo{author}{\bibfnamefont{M.}~\bibnamefont{Kaplinghat}},
  \bibnamefont{and} \bibinfo{author}{\bibfnamefont{H.-B.} \bibnamefont{Yu}},
  \bibinfo{journal}{Phys. Rev. Lett.} \textbf{\bibinfo{volume}{104}},
  \bibinfo{pages}{151301} (\bibinfo{year}{2010}), \eprint{0911.0422}.

\bibitem[{\citenamefont{Buckley and Fox}(2010)}]{Buckley:2009in}
\bibinfo{author}{\bibfnamefont{M.~R.} \bibnamefont{Buckley}} \bibnamefont{and}
  \bibinfo{author}{\bibfnamefont{P.~J.} \bibnamefont{Fox}},
  \bibinfo{journal}{Phys. Rev.} \textbf{\bibinfo{volume}{D81}},
  \bibinfo{pages}{083522} (\bibinfo{year}{2010}), \eprint{0911.3898}.

\bibitem[{\citenamefont{Bjorken et~al.}(2009)\citenamefont{Bjorken, Essig,
  Schuster, and Toro}}]{Bjorken:2009mm}
\bibinfo{author}{\bibfnamefont{J.~D.} \bibnamefont{Bjorken}},
  \bibinfo{author}{\bibfnamefont{R.}~\bibnamefont{Essig}},
  \bibinfo{author}{\bibfnamefont{P.}~\bibnamefont{Schuster}}, \bibnamefont{and}
  \bibinfo{author}{\bibfnamefont{N.}~\bibnamefont{Toro}},
  \bibinfo{journal}{Phys. Rev.} \textbf{\bibinfo{volume}{D80}},
  \bibinfo{pages}{075018} (\bibinfo{year}{2009}), \eprint{0906.0580}.

\bibitem[{\citenamefont{Schuster et~al.}(2010)\citenamefont{Schuster, Toro, and
  Yavin}}]{Schuster:2009au}
\bibinfo{author}{\bibfnamefont{P.}~\bibnamefont{Schuster}},
  \bibinfo{author}{\bibfnamefont{N.}~\bibnamefont{Toro}}, \bibnamefont{and}
  \bibinfo{author}{\bibfnamefont{I.}~\bibnamefont{Yavin}},
  \bibinfo{journal}{Phys. Rev.} \textbf{\bibinfo{volume}{D81}},
  \bibinfo{pages}{016002} (\bibinfo{year}{2010}), \eprint{0910.1602}.

\bibitem[{\citenamefont{Essig et~al.}(2010{\natexlab{a}})\citenamefont{Essig,
  Schuster, Toro, and Wojtsekhowski}}]{Essig:2010xa}
\bibinfo{author}{\bibfnamefont{R.}~\bibnamefont{Essig}},
  \bibinfo{author}{\bibfnamefont{P.}~\bibnamefont{Schuster}},
  \bibinfo{author}{\bibfnamefont{N.}~\bibnamefont{Toro}}, \bibnamefont{and}
  \bibinfo{author}{\bibfnamefont{B.}~\bibnamefont{Wojtsekhowski}}
  (\bibinfo{year}{2010}{\natexlab{a}}), \eprint{1001.2557}.

\bibitem[{\citenamefont{Khrapak et~al.}(2003)\citenamefont{Khrapak, Ivlev,
  Morfill, and Zhdanov}}]{PhysRevLett.90.225002}
\bibinfo{author}{\bibfnamefont{S.~A.} \bibnamefont{Khrapak}},
  \bibinfo{author}{\bibfnamefont{A.~V.} \bibnamefont{Ivlev}},
  \bibinfo{author}{\bibfnamefont{G.~E.} \bibnamefont{Morfill}},
  \bibnamefont{and} \bibinfo{author}{\bibfnamefont{S.~K.}
  \bibnamefont{Zhdanov}}, \bibinfo{journal}{Phys. Rev. Lett.}
  \textbf{\bibinfo{volume}{90}}, \bibinfo{pages}{225002}
  (\bibinfo{year}{2003}).

\bibitem[{\citenamefont{Finkbeiner et~al.}(2010)\citenamefont{Finkbeiner,
  Goodenough, Slatyer, Vogelsberger, and Weiner}}]{Finkbeiner:2010sm}
\bibinfo{author}{\bibfnamefont{D.~P.} \bibnamefont{Finkbeiner}},
  \bibinfo{author}{\bibfnamefont{L.}~\bibnamefont{Goodenough}},
  \bibinfo{author}{\bibfnamefont{T.~R.} \bibnamefont{Slatyer}},
  \bibinfo{author}{\bibfnamefont{M.}~\bibnamefont{Vogelsberger}},
  \bibnamefont{and} \bibinfo{author}{\bibfnamefont{N.}~\bibnamefont{Weiner}}
  (\bibinfo{year}{2010}), \eprint{1011.3082}.

\bibitem[{\citenamefont{de~Naray et~al.}(2009)\citenamefont{de~Naray, Martinez,
  Bullock, and Kaplinghat}}]{Naray}
\bibinfo{author}{\bibfnamefont{R.~K.} \bibnamefont{de~Naray}},
  \bibinfo{author}{\bibfnamefont{G.~D.} \bibnamefont{Martinez}},
  \bibinfo{author}{\bibfnamefont{J.~S.} \bibnamefont{Bullock}},
  \bibnamefont{and}
  \bibinfo{author}{\bibfnamefont{M.}~\bibnamefont{Kaplinghat}}
  (\bibinfo{year}{2009}), \eprint{0912.3518}.

\bibitem[{\citenamefont{Gnedin and Ostriker}(2000)}]{Gnedin:2000ea}
\bibinfo{author}{\bibfnamefont{O.~Y.} \bibnamefont{Gnedin}} \bibnamefont{and}
  \bibinfo{author}{\bibfnamefont{J.~P.} \bibnamefont{Ostriker}},
  \bibinfo{journal}{Astrophys.J.}  (\bibinfo{year}{2000}),
  \eprint{astro-ph/0010436}.

\bibitem[{\citenamefont{Tucker-Smith and Weiner}(2001)}]{TuckerSmith:2001hy}
\bibinfo{author}{\bibfnamefont{D.}~\bibnamefont{Tucker-Smith}}
  \bibnamefont{and} \bibinfo{author}{\bibfnamefont{N.}~\bibnamefont{Weiner}},
  \bibinfo{journal}{Phys. Rev.} \textbf{\bibinfo{volume}{D64}},
  \bibinfo{pages}{043502} (\bibinfo{year}{2001}), \eprint{hep-ph/0101138}.

\bibitem[{\citenamefont{Finkbeiner et~al.}(2009)\citenamefont{Finkbeiner,
  Slatyer, Weiner, and Yavin}}]{Finkbeiner:2009mi}
\bibinfo{author}{\bibfnamefont{D.~P.} \bibnamefont{Finkbeiner}},
  \bibinfo{author}{\bibfnamefont{T.~R.} \bibnamefont{Slatyer}},
  \bibinfo{author}{\bibfnamefont{N.}~\bibnamefont{Weiner}}, \bibnamefont{and}
  \bibinfo{author}{\bibfnamefont{I.}~\bibnamefont{Yavin}},
  \bibinfo{journal}{JCAP} \textbf{\bibinfo{volume}{0909}}, \bibinfo{pages}{037}
  (\bibinfo{year}{2009}), \eprint{0903.1037}.

\bibitem[{\citenamefont{Batell et~al.}(2009)\citenamefont{Batell, Pospelov, and
  Ritz}}]{Batell:2009vb}
\bibinfo{author}{\bibfnamefont{B.}~\bibnamefont{Batell}},
  \bibinfo{author}{\bibfnamefont{M.}~\bibnamefont{Pospelov}}, \bibnamefont{and}
  \bibinfo{author}{\bibfnamefont{A.}~\bibnamefont{Ritz}},
  \bibinfo{journal}{Phys.Rev.} \textbf{\bibinfo{volume}{D79}},
  \bibinfo{pages}{115019} (\bibinfo{year}{2009}), \eprint{0903.3396}.

\bibitem[{\citenamefont{Lang and Weiner}(2010)}]{Lang:2010cd}
\bibinfo{author}{\bibfnamefont{R.~F.} \bibnamefont{Lang}} \bibnamefont{and}
  \bibinfo{author}{\bibfnamefont{N.}~\bibnamefont{Weiner}},
  \bibinfo{journal}{JCAP} \textbf{\bibinfo{volume}{1006}}, \bibinfo{pages}{032}
  (\bibinfo{year}{2010}), \eprint{1003.3664}.

\bibitem[{\citenamefont{Essig et~al.}(2010{\natexlab{b}})\citenamefont{Essig,
  Kaplan, Schuster, and Toro}}]{Essig:2010ye}
\bibinfo{author}{\bibfnamefont{R.}~\bibnamefont{Essig}},
  \bibinfo{author}{\bibfnamefont{J.}~\bibnamefont{Kaplan}},
  \bibinfo{author}{\bibfnamefont{P.}~\bibnamefont{Schuster}}, \bibnamefont{and}
  \bibinfo{author}{\bibfnamefont{N.}~\bibnamefont{Toro}},
  \bibinfo{journal}{Submitted to Physical Review D}
  (\bibinfo{year}{2010}{\natexlab{b}}), \eprint{1004.0691}.

\bibitem[{\citenamefont{Graham et~al.}(2010)\citenamefont{Graham, Harnik,
  Rajendran, and Saraswat}}]{Graham:2010ca}
\bibinfo{author}{\bibfnamefont{P.~W.} \bibnamefont{Graham}},
  \bibinfo{author}{\bibfnamefont{R.}~\bibnamefont{Harnik}},
  \bibinfo{author}{\bibfnamefont{S.}~\bibnamefont{Rajendran}},
  \bibnamefont{and} \bibinfo{author}{\bibfnamefont{P.}~\bibnamefont{Saraswat}},
  \bibinfo{journal}{Phys.Rev.} \textbf{\bibinfo{volume}{D82}},
  \bibinfo{pages}{063512} (\bibinfo{year}{2010}), \eprint{1004.0937}.

\bibitem[{\citenamefont{Finkbeiner and Weiner}(2007)}]{Finkbeiner:2007kk}
\bibinfo{author}{\bibfnamefont{D.~P.} \bibnamefont{Finkbeiner}}
  \bibnamefont{and} \bibinfo{author}{\bibfnamefont{N.}~\bibnamefont{Weiner}},
  \bibinfo{journal}{Phys. Rev.} \textbf{\bibinfo{volume}{D76}},
  \bibinfo{pages}{083519} (\bibinfo{year}{2007}), \eprint{astro-ph/0702587}.

\bibitem[{\citenamefont{Pospelov and Ritz}(2007)}]{Pospelov:2007xh}
\bibinfo{author}{\bibfnamefont{M.}~\bibnamefont{Pospelov}} \bibnamefont{and}
  \bibinfo{author}{\bibfnamefont{A.}~\bibnamefont{Ritz}},
  \bibinfo{journal}{Phys.Lett.} \textbf{\bibinfo{volume}{B651}},
  \bibinfo{pages}{208} (\bibinfo{year}{2007}), \eprint{hep-ph/0703128}.

\bibitem[{\citenamefont{Chen et~al.}(2009)\citenamefont{Chen, Cline, and
  Frey}}]{Chen:2009dm}
\bibinfo{author}{\bibfnamefont{F.}~\bibnamefont{Chen}},
  \bibinfo{author}{\bibfnamefont{J.~M.} \bibnamefont{Cline}}, \bibnamefont{and}
  \bibinfo{author}{\bibfnamefont{A.~R.} \bibnamefont{Frey}},
  \bibinfo{journal}{Phys.Rev.} \textbf{\bibinfo{volume}{D79}},
  \bibinfo{pages}{063530} (\bibinfo{year}{2009}), \eprint{0901.4327}.

\bibitem[{\citenamefont{Chen et~al.}(2010)\citenamefont{Chen, Cline, Fradette,
  Frey, and Rabideau}}]{Chen:2009av}
\bibinfo{author}{\bibfnamefont{F.}~\bibnamefont{Chen}},
  \bibinfo{author}{\bibfnamefont{J.~M.} \bibnamefont{Cline}},
  \bibinfo{author}{\bibfnamefont{A.}~\bibnamefont{Fradette}},
  \bibinfo{author}{\bibfnamefont{A.~R.} \bibnamefont{Frey}}, \bibnamefont{and}
  \bibinfo{author}{\bibfnamefont{C.}~\bibnamefont{Rabideau}},
  \bibinfo{journal}{Phys.Rev.} \textbf{\bibinfo{volume}{D81}},
  \bibinfo{pages}{043523} (\bibinfo{year}{2010}), \eprint{0911.2222}.

\bibitem[{\citenamefont{Goodman and Hut}(1993)}]{Goodman}
\bibinfo{author}{\bibfnamefont{J.}~\bibnamefont{Goodman}} \bibnamefont{and}
  \bibinfo{author}{\bibfnamefont{P.}~\bibnamefont{Hut}},
  \bibinfo{journal}{Astrophys. J.} \textbf{\bibinfo{volume}{403}},
  \bibinfo{pages}{271} (\bibinfo{year}{1993}).

\bibitem[{\citenamefont{Strigari et~al.}(2008)}]{Strigari}
\bibinfo{author}{\bibfnamefont{L.~E.} \bibnamefont{Strigari}}
  \bibnamefont{et~al.}, \bibinfo{journal}{Nature}
  \textbf{\bibinfo{volume}{454}}, \bibinfo{pages}{1096} (\bibinfo{year}{2008}),
  \eprint{0808.3772}.

\bibitem[{\citenamefont{Bullock}(2010)}]{Bullock}
\bibinfo{author}{\bibfnamefont{J.~S.} \bibnamefont{Bullock}}
  (\bibinfo{year}{2010}), \eprint{1009.4505}.

\bibitem[{\citenamefont{Conroy et~al.}(2010)\citenamefont{Conroy, Loeb, and
  Spergel}}]{Conroy}
\bibinfo{author}{\bibfnamefont{C.}~\bibnamefont{Conroy}},
  \bibinfo{author}{\bibfnamefont{A.}~\bibnamefont{Loeb}}, \bibnamefont{and}
  \bibinfo{author}{\bibfnamefont{D.}~\bibnamefont{Spergel}}
  (\bibinfo{year}{2010}), \eprint{1010.5783}.

\bibitem[{\citenamefont{Loeb}(2010)}]{Book}
\bibinfo{author}{\bibfnamefont{A.}~\bibnamefont{Loeb}},
  \emph{\bibinfo{title}{How Did The First Stars and Galaxies Form?}}
  (\bibinfo{publisher}{Princeton Univ. Press}, \bibinfo{year}{2010}).

\bibitem[{\citenamefont{Pritchard and Loeb}(2010)}]{Pritchard}
\bibinfo{author}{\bibfnamefont{J.~R.} \bibnamefont{Pritchard}}
  \bibnamefont{and} \bibinfo{author}{\bibfnamefont{A.}~\bibnamefont{Loeb}},
  \bibinfo{journal}{Phys. Rev.} \textbf{\bibinfo{volume}{D82}},
  \bibinfo{pages}{023006} (\bibinfo{year}{2010}), \eprint{1005.4057}.

\bibitem[{\citenamefont{{Pritchard} et~al.}(2010)\citenamefont{{Pritchard},
  {Loeb}, and {Wyithe}}}]{2010MNRAS.408...57P}
\bibinfo{author}{\bibfnamefont{J.~R.} \bibnamefont{{Pritchard}}},
  \bibinfo{author}{\bibfnamefont{A.}~\bibnamefont{{Loeb}}}, \bibnamefont{and}
  \bibinfo{author}{\bibfnamefont{J.~S.~B.} \bibnamefont{{Wyithe}}},
  \bibinfo{journal}{\mnras} \textbf{\bibinfo{volume}{408}}, \bibinfo{pages}{57}
  (\bibinfo{year}{2010}), \eprint{0908.3891}.

\bibitem[{\citenamefont{Baumgart et~al.}(2009)\citenamefont{Baumgart, Cheung,
  Ruderman, Wang, and Yavin}}]{Baumgart:2009tn}
\bibinfo{author}{\bibfnamefont{M.}~\bibnamefont{Baumgart}},
  \bibinfo{author}{\bibfnamefont{C.}~\bibnamefont{Cheung}},
  \bibinfo{author}{\bibfnamefont{J.~T.} \bibnamefont{Ruderman}},
  \bibinfo{author}{\bibfnamefont{L.-T.} \bibnamefont{Wang}}, \bibnamefont{and}
  \bibinfo{author}{\bibfnamefont{I.}~\bibnamefont{Yavin}},
  \bibinfo{journal}{JHEP} \textbf{\bibinfo{volume}{04}}, \bibinfo{pages}{014}
  (\bibinfo{year}{2009}), \eprint{0901.0283}.

\bibitem[{\citenamefont{Kaplan}(1992)}]{Kaplan:1991ah}
\bibinfo{author}{\bibfnamefont{D.~B.} \bibnamefont{Kaplan}},
  \bibinfo{journal}{Phys. Rev. Lett.} \textbf{\bibinfo{volume}{68}},
  \bibinfo{pages}{741} (\bibinfo{year}{1992}).

\bibitem[{\citenamefont{Kitano and Low}(2005)}]{Kitano:2004sv}
\bibinfo{author}{\bibfnamefont{R.}~\bibnamefont{Kitano}} \bibnamefont{and}
  \bibinfo{author}{\bibfnamefont{I.}~\bibnamefont{Low}},
  \bibinfo{journal}{Phys. Rev.} \textbf{\bibinfo{volume}{D71}},
  \bibinfo{pages}{023510} (\bibinfo{year}{2005}), \eprint{hep-ph/0411133}.

\bibitem[{\citenamefont{Farrar and Zaharijas}(2006)}]{Farrar:2005zd}
\bibinfo{author}{\bibfnamefont{G.~R.} \bibnamefont{Farrar}} \bibnamefont{and}
  \bibinfo{author}{\bibfnamefont{G.}~\bibnamefont{Zaharijas}},
  \bibinfo{journal}{Phys. Rev. Lett.} \textbf{\bibinfo{volume}{96}},
  \bibinfo{pages}{041302} (\bibinfo{year}{2006}), \eprint{hep-ph/0510079}.

\bibitem[{\citenamefont{Kaplan et~al.}(2009)\citenamefont{Kaplan, Luty, and
  Zurek}}]{Kaplan:2009ag}
\bibinfo{author}{\bibfnamefont{D.~E.} \bibnamefont{Kaplan}},
  \bibinfo{author}{\bibfnamefont{M.~A.} \bibnamefont{Luty}}, \bibnamefont{and}
  \bibinfo{author}{\bibfnamefont{K.~M.} \bibnamefont{Zurek}},
  \bibinfo{journal}{Phys. Rev.} \textbf{\bibinfo{volume}{D79}},
  \bibinfo{pages}{115016} (\bibinfo{year}{2009}), \eprint{0901.4117}.

\bibitem[{\citenamefont{Chang and Luty}(2009)}]{Chang:2009sv}
\bibinfo{author}{\bibfnamefont{S.}~\bibnamefont{Chang}} \bibnamefont{and}
  \bibinfo{author}{\bibfnamefont{M.~A.} \bibnamefont{Luty}}
  (\bibinfo{year}{2009}), \eprint{0906.5013}.

\bibitem[{\citenamefont{Cohen et~al.}(2010)\citenamefont{Cohen, Phalen, Pierce,
  and Zurek}}]{Cohen:2010kn}
\bibinfo{author}{\bibfnamefont{T.}~\bibnamefont{Cohen}},
  \bibinfo{author}{\bibfnamefont{D.~J.} \bibnamefont{Phalen}},
  \bibinfo{author}{\bibfnamefont{A.}~\bibnamefont{Pierce}}, \bibnamefont{and}
  \bibinfo{author}{\bibfnamefont{K.~M.} \bibnamefont{Zurek}},
  \bibinfo{journal}{Phys. Rev.} \textbf{\bibinfo{volume}{D82}},
  \bibinfo{pages}{056001} (\bibinfo{year}{2010}), \eprint{1005.1655}.

\end{thebibliography}
\bibliographystyle{apsrev}

\end{document}